\documentclass[aps,showkeys,showpacs,twocolumn,showpacs,preprintnumbers,amsmath,amssymb,superscriptaddress,floatfix,nofootinbib]{revtex4}
\usepackage{graphicx,color,dcolumn,booktabs,bm}
\usepackage{longtable,lscape}
\usepackage{txfonts}
\usepackage{overpic}
\usepackage{epsfig}
\usepackage{amssymb}
\usepackage{rotating}
\usepackage{epstopdf}
\usepackage{indentfirst}
\usepackage{feynmf}   %{feynmp}
\usepackage{slashed}  %for Feynman symbols
\usepackage{cases}
\usepackage{color}
\usepackage{multirow}
\usepackage{graphicx,color,dcolumn,booktabs,bm}
\usepackage{cases}
\usepackage{array}

\graphicspath{{Figures/}} %
\usepackage[colorlinks, citecolor=blue,anchorcolor=red,menucolor=red, linkcolor=red,filecolor=red,runcolor=red,urlcolor=blue,frenchlinks=red]{hyperref}

\newcommand{\vsig}{\mbox{\boldmath$\sigma$\unboldmath}}
\newcommand{\veps}{\mbox{\boldmath$\epsilon$\unboldmath}}
\newcommand{\vrho}{\mbox{\boldmath$\rho$\unboldmath}}
\newcommand{\vlab}{\mbox{\boldmath$\lambda$\unboldmath}}

\begin{document}

\title{Triply charmed and bottom baryons in a constituent quark model}

\author{Ming-Sheng Liu} \email{liumingsheng0001@126.com}

\affiliation{  Department
of Physics, Hunan Normal University,  Changsha 410081, China }

\affiliation{ Synergetic Innovation
Center for Quantum Effects and Applications (SICQEA), Changsha 410081,China}

\affiliation{  Key Laboratory of
Low-Dimensional Quantum Structures and Quantum Control of Ministry
of Education, Changsha 410081, China}

\author{Qi-Fang L\"{u}} \email{lvqifang@hunnu.edu.cn}
\affiliation{  Department
of Physics, Hunan Normal University,  Changsha 410081, China }

\affiliation{ Synergetic Innovation
Center for Quantum Effects and Applications (SICQEA), Changsha 410081,China}

\affiliation{  Key Laboratory of
Low-Dimensional Quantum Structures and Quantum Control of Ministry
of Education, Changsha 410081, China}

\author{Xian-Hui Zhong \footnote{Corresponding author} } \email{zhongxh@hunnu.edu.cn} %
\affiliation{  Department
of Physics, Hunan Normal University,  Changsha 410081, China }

\affiliation{ Synergetic Innovation
Center for Quantum Effects and Applications (SICQEA), Changsha 410081,China}

\affiliation{  Key Laboratory of
Low-Dimensional Quantum Structures and Quantum Control of Ministry
of Education, Changsha 410081, China}

\begin{abstract}
In this work, we study the mass spectrum of the $\Omega_{ccc}$ and $\Omega_{bbb}$ baryons up to the $N=2$ shell within a nonrelativistic constituent quark model (NRCQM). The model parameters are adopted from the determinations by fitting the charmonium and bottomonium
spectra in our previous works. The masses of the $\Omega_{ccc}$ and $\Omega_{bbb}$ baryon states predicted in present work reasonably agree with
the results obtained with the Lattice QCD calculations. Furthermore, to provide more knowledge of the $\Omega_{ccc}$ and $\Omega_{bbb}$ states,
we evaluate their radiative decays with the available masses and wave functions from the potential model.
\end{abstract}

\maketitle

\section{Introduction}{\label{introduction}}

In the discovery of the heavy baryons, the Large Hadron Collider (LHC) facility has shown its powerful abilities in recent years. For example, in 2017 the first doubly charmed baryon $\Xi_{cc}^{++}(3621)$ was discovered in the $\Lambda_c^+K^-\pi^+\pi^-$ mass spectrum~\cite{Aaij:2017ueg}, and was confirmed in the $\Xi_c^+\pi^+$ channel one year later~\cite{Aaij:2018gfl} by the LHCb Collaboration. In 2017, five extremely narrow $\Omega_c(X)$ states, $\Omega_c(3000)$, $\Omega_c(3050)$, $\Omega_c(3066)$, $\Omega_c(3090)$ and $\Omega_c(3119)$, were observed in the $\Xi_c^{+}K^-$ channel by the LHCb Collaboration~\cite{Aaij:2017nav}.
Very recently four bottom baryon resonances  $\Xi_b(6227)^-$~\cite{Aaij:2018yqz}, $\Sigma_b(6097)^{\pm}$~\cite{Aaij:2018tnn}, $\Lambda_b(6146/6152)^0$~\cite{Aaij:2019amv} were observed at LHCb as well. Except for the singly and doubly heavy baryons,
the LHC facility may provide good opportunities for discovering the missing triply heavy baryons~\cite{production1,production3}.

The triply heavy baryons, as a system of fully heavy quarks, may provide a new window for understanding the structure of baryons.
The complications of light-quark interaction are absent in the triply heavy baryons, thus, they provide an ideal place for
our better understanding the heavy quark dynamics. From the theoretical point view, the potential models
might be able to describe triply heavy baryons to a similar level of precision as their success in heavy quarkonia.
Just as the quark-antiquark interactions are examined in charmonia and bottomonia, the studies of the triply heavy baryon spectra will probe the quark-quark interactions in the heavy quark sector~\cite{ref17}. In the past years, many studies about
the triply heavy baryons can be found in the literature. Most of them focused on the predictions of the masses~\cite{ref1,ref2,ref3,ref5,ref6,ref7,ref8,ref9,ref10,ref11,ref12,ref13,ref14,ref15,ref16,
ref17,ref18,ref19,ref20,ref21,ref22,ref23,ref24,ref25,ref26,ref27,ref28,ref29,ref30,ref31} and the production~\cite{production1,production2,production3,production4,production5,production6,production7,
production8}. However, only several works have paid attentions to the weak
decays~\cite{ref4,Wang:2018utj,production11}, magnetic moments~\cite{ref22,Simonis:2018rld}, $M1$ decays~\cite{Simonis:2018rld}
of triply heavy baryons.

Stimulated by the large discovery potentials of the heavy baryons at the LHC facility, in this work
we carry out a systemical study of the triply heavy baryon spectra of $\Omega_{ccc}$ and $\Omega_{bbb}$ within a
nonrelativistic potential model. Recently, with this model we have studied the spectra of the charmonium,
bottomonium, $B_c$ meson, $\Omega$ baryon, and fully-heavy tetraquark states. For there are
no measurements of the triply heavy baryons which can be used to constrain the parameters potential model,
the model parameters of this work are adopted with the determinations by fitting the charmonium and bottomonium
spectra in our previous works~\cite{Deng:2016stx, Liu:2019zuc, Li:2019qsg}. The masses of the $\Omega_{ccc}$ and
$\Omega_{bbb}$ baryon states predicted in present work reasonably agree with
the results obtained with the Lattice QCD calculations~\cite{ref17,ref18}.

Furthermore, to provide more knowledge of the $\Omega_{ccc}$ and $\Omega_{bbb}$ states,
we evaluate their radiative decays with the available wave functions from the potential model.
It should be emphasized that the OZI allowed two-body strong decay channels are absence
for the low-lying $1P$-, $1D$-, and $2S$-wave $\Omega_{ccc}$ and $\Omega_{bbb}$ states, thus,
the radiative transitions become important in their decays. Consequently, the radiative decay processes of the excited $\Omega_{ccc}$ or $\Omega_{bbb}$ states may be crucial for establishing them if they are produced in experiments. In this work, radiative decays of
the $\Omega_{ccc}$ and $\Omega_{bbb}$ states are calculated within a nonrelativistic constituent
quark model developed in our previous study of the heavy quarkonia~\cite{Deng:2016stx,Deng:2016ktl}.
This model was also successfully extended to deal with the radiative decays of the $B_c$ meson states~\cite{Li:2019tbn}, $\Omega$ baryon states~\cite{LZ}, singly baryon states~\cite{Yao:2018jmc,Wang:2017kfr,Wang:2017hej}, and doubly heavy baryon states~\cite{Xiao:2017udy,Lu:2017meb}.

This paper is organized as follows. In Sec.~\ref{Spectra}, a brief review of the potential model is given and the mass spectra of the $\Omega_{ccc}$ and $\Omega_{bbb}$ baryons are calculated. Then, in Sec.~\ref{radiative decay},
we give a review of the radiative decay model, and calculate radiative decays
of the excited $\Omega_{ccc}$ and $\Omega_{bbb}$ states by using the masses and wave functions obtained from the potential model.
In Sec.~\ref{discussions}, we give our discussions based on the obtained radiative decay properties and masses of the $\Omega_{ccc}$ and $\Omega_{bbb}$ resonances. Finally, a summary is given in Sec.~\ref{Summary}.

\section{Mass spectrum} \label{Spectra}

\subsection{Hamiltonian} \label{Hamiltonian}

To calculate the spectrum of the $\Omega_{ccc}$ and $\Omega_{bbb}$ baryons, the following nonrelativistic Hamiltonian is adopted in this work
\begin{equation}\label{Hamiltonian}
H=(\sum_{i=1}^3 m_i+T_i)-T_G+\sum_{i<j}V_{ij}(r_{ij}),
\end{equation}
where $m_i$ and $T_i$ stand for the constituent quark mass and kinetic energy of the $i$-th quark, respectively; $T_G$ stands for the center-of-mass (c.m.) kinetic energy of the baryon system; $r_{ij}\equiv|\mathbf{r}_i-\mathbf{r}_j|$ is the distance between the $i$-th quark and  $j$-th quark; and $V_{ij}(r_{ij})$ stands for the effective potential between the $i$-th and  $j$-th quark.
In this work, we adopt a widely used potential form for $V_{ij}(r_{ij})$~\cite{Eichten:1978tg,Godfrey:1985xj,Swanson:2005,Godfrey:2015dia,Godfrey:2004ya,Lakhina:2006fy,Lu:2016bbk,Li:2010vx,Deng:2016stx,Deng:2016ktl}, i.e.
\begin{equation}\label{vij}
V_{ij}(r_{ij})=V_{ij}^{conf}(r_{ij})+V_{ij}^{sd}(r_{ij}) \ ,
\end{equation}
where $V^{conf}_{ij}$ stands for the potential for confinement, and is adopted the standard Cornell form:
\begin{equation}\label{vcon}
V_{ij}^{conf}(r_{ij})=\frac{b}{2}r_{ij}-\frac{2}{3}\frac{\alpha_{ij}}{r_{ij}},
\end{equation}
while $V_{ij}^{sd}(r_{ij})$ stands for the spin-dependent interaction, which is the sum of the spin-spin contact hyperfine potential $V_{ij}^{SS}$,
the tensor term $V_{ij}^{T}$, and the spin-orbit term $V_{ij}^{LS}$:
\begin{equation}\label{voge cen}
V^{sd}_{ij}=V^{SS}_{ij}+V^{T}_{ij}+V^{LS}_{ij}.
\end{equation}
The spin-spin potential $V_{ij}^{SS}$ and the tensor term $V_{ij}^{T}$ are adopted the often used forms:
\begin{equation}\label{voge cen}
V^{SS}_{ij}=-\frac{2\alpha_{ij}}{3}\left\{-\frac{\pi}{2}\cdot\frac{\sigma^3_{ij}e^{-\sigma^2_{ij}r_{ij}^2}}{\pi^{3/2}}\cdot\frac{16}{3m_im_j}(\mathbf{S}_i\cdot\mathbf{S}_j)\right\},
\end{equation}
\begin{equation}\label{voge ten}
V^{T}_{ij}=\frac{2\alpha_{ij}}{3}\cdot\frac{1}{m_im_jr_{ij}^3}\Bigg\{\frac{3(\mathbf{S}_i\cdot \mathbf{r}_{ij})(\mathbf{S}_j\cdot \mathbf{r}_{ij})}{r_{ij}^2}-\mathbf{S}_i\cdot\mathbf{S}_j\Bigg\}.
\end{equation}
In this work, a simplified phenomenological spin-orbit potential is adopted as that suggested in the literature~\cite{Pervin:2007wa,Roberts:2007ni,LZ}, i.e.,
\begin{equation}\label{voge LS}
V^{LS}_{ij}=\frac{\alpha_{\mathrm{SO}}}{\rho^2+\lambda^2}\cdot \frac{\mathbf{L}\cdot\mathbf{S}}{3(m_1+m_2+m_3)^2}.
\end{equation}
In the above equations, the $\mathbf{S}_i$, $\mathbf{S}$ and $\mathbf{L}$ are the spin operator of the $i$-th quark, the total spin of the baryon and the total orbital angular momentum of the baryon, respectively;
the parameter $b$, $\alpha_{ij}$, and $\alpha_{SO}$ denote the strength of confinement potential, strong coupling, and spin-orbit potential, respectively.

The seven parameters $m_c$, $m_b$, ${\alpha_{cc}}$, ${\alpha_{bb}}$, ${\sigma_{cc}}$, ${\sigma_{bb}}$, and $b$ have been determined by fitting the charmonium and bottomonium spectra in our previous works~\cite{Deng:2016stx, Liu:2019zuc, Li:2019qsg}. In this work, we use the same value of parameter $\alpha_{SO}$ as in Ref.~\cite{LZ}. The quark model parameters adopted in present work are collected in Table~\ref{Quark model parameters}.

\begin{table}[htp]
\begin{center}
\caption{\label{Quark model parameters} Quark model parameters adopted in this work.}
\scalebox{1.0}{
\begin{tabular}{lllllllllllllllll}\hline\hline
~~~~~~~~~~~~&Parameter &~~~~~~~~~~~~~~~~~~~~~~~~~~~~~~~~~~~~~~Value~~~~~~~~~~~~~~~~~~~\\
\hline
~~~~~~~~~~~~&$m_c$             ~$(\textrm{GeV})$       &~~~~~~~~~~~~~~~~~~~~~~~~~~~~~~~~~~~~~~1.4830~~~~~~~~~~~~~~~~~~~\\
~~~~~~~~~~~~&$m_b$             ~$(\textrm{GeV})$       &~~~~~~~~~~~~~~~~~~~~~~~~~~~~~~~~~~~~~~4.8520~~~~~~~~~~~~~~~~~~~\\
~~~~~~~~~~~~&${\alpha_{cc}}$                           &~~~~~~~~~~~~~~~~~~~~~~~~~~~~~~~~~~~~~~0.5461~~~~~~~~~~~~~~~~~~~\\
~~~~~~~~~~~~&${\alpha_{bb}}$                           &~~~~~~~~~~~~~~~~~~~~~~~~~~~~~~~~~~~~~~0.4311~~~~~~~~~~~~~~~~~~~\\
~~~~~~~~~~~~&${\sigma_{cc}}$   ~$(\textrm{GeV})$       &~~~~~~~~~~~~~~~~~~~~~~~~~~~~~~~~~~~~~~1.1384~~~~~~~~~~~~~~~~~~~\\
~~~~~~~~~~~~&${\sigma_{bb}}$   ~$(\textrm{GeV})$       &~~~~~~~~~~~~~~~~~~~~~~~~~~~~~~~~~~~~~~2.3200~~~~~~~~~~~~~~~~~~~\\
~~~~~~~~~~~~&$b$               ~$(\textrm{GeV}^2)$     &~~~~~~~~~~~~~~~~~~~~~~~~~~~~~~~~~~~~~~0.1425~~~~~~~~~~~~~~~~~~~\\
~~~~~~~~~~~~&${\alpha_{SO}}$   ~$(\textrm{GeV})$       &~~~~~~~~~~~~~~~~~~~~~~~~~~~~~~~~~~~~~~1.9000~~~~~~~~~~~~~~~~~~~\\
\hline\hline
\end{tabular}}
\end{center}
\end{table}

\begin{table*}[htp]
\begin{center}
\caption{\label{wave function table} Quark model classification of the $\Omega_{QQQ}$ ($Q=s,c,b$) states up to $N=2$ shell. }
\scalebox{1.0}{
\begin{tabular}{cccccccccccc}\hline\hline
  &$n^{2S+1}L_{J^P}$          ~~~~~~&$|N_6, ^{2S+1}N_3, N, L, J^P\rangle$~~~~~~& SU(6)$\times$O(3) wave functions \\
\hline
  &$1^{4}S_{\frac{3}{2}^{+}}$ ~~~~~~&$|56, ^{4}10, 0, 0, \frac{3}{2}^{+}\rangle$~~~~~~&$\psi^{S}_{000}(\vrho, \vlab)\chi^{S}_{M_S}$~~~~~~\\

  &$1^{2}P_{\frac{1}{2}^{-}}$~~~~~~&$|70, ^{2}10, 1, 1, \frac{1}{2}^{-}\rangle$
~~~~~~& $\sum\limits_{M_L+M_S=M} \langle 1 M_L; \frac{1}{2} M_S|\frac{1}{2} M\rangle \sqrt{\frac{1}{2}}\bigg(\psi^{\lambda}_{11M_L}(\vrho, \vlab)\chi^{\lambda}_{M_S}+\psi^{\rho}_{11M_L}(\vrho, \vlab)\chi^{\rho}_{M_S}\bigg)$ ~~~~~~\\
& $1^{2}P_{\frac{3}{2}^{-}}$ ~~~~~~& $|70, ^{2}10, 1, 1, \frac{3}{2}^{-}\rangle$
~~~~~~& $\sum\limits_{M_L+M_S=M} \langle 1 M_L; \frac{1}{2} M_S|\frac{3}{2} M\rangle \sqrt{\frac{1}{2}}\bigg(\psi^{\lambda}_{11M_L}(\vrho, \vlab)\chi^{\lambda}_{M_S}+\psi^{\rho}_{11M_L}(\vrho, \vlab)\chi^{\rho}_{M_S}\bigg)$~~~~~~\\
& $2^{2}S_{\frac{1}{2}^{+}}$  ~~~~~~& $|70, ^{2}10, 2, 0, \frac{1}{2}^{+}\rangle$
~~~~~~& $\sqrt{\frac{1}{2}}\bigg(\psi^{\lambda}_{200}(\vrho, \vlab)\chi^{\lambda}_{M_S}+\psi^{\rho}_{200}(\vrho, \vlab)\chi^{\rho}_{M_S}\bigg)$~~~~~~\\
& $2^{4}S_{\frac{3}{2}^{+}}$  ~~~~~~& $|56, ^{4}10, 2, 0, \frac{3}{2}^{+}\rangle$
~~~~~~& $\psi^{S}_{200}(\vrho, \vlab)\chi^{S}_{M_S}$ ~~~~~~\\
&$1^{2}D_{\frac{3}{2}^{+}}$~~~~~~&$|70, ^{2}10, 2, 2, \frac{3}{2}^{+}\rangle$
~~~~~~&$\sum\limits_{M_L+M_S=M} \langle 2 M_L; \frac{1}{2} M_S|\frac{3}{2} M\rangle \sqrt{\frac{1}{2}}\bigg(\psi^{\lambda}_{22M_L}(\vrho, \vlab)\chi^{\lambda}_{M_S}+\psi^{\rho}_{22M_L}(\vrho, \vlab)\chi^{\rho}_{M_S}\bigg)$
~~~~~~\\
&$1^{2}D_{\frac{5}{2}^{+}}$ ~~~~~~&$|70, ^{2}10, 2, 2, \frac{5}{2}^{+}\rangle$
~~~~~~&$\sum\limits_{M_L+M_S=M} \langle 2 M_L; \frac{1}{2} M_S|\frac{5}{2} M\rangle \sqrt{\frac{1}{2}}\bigg(\psi^{\lambda}_{22M_L}(\vrho, \vlab)\chi^{\lambda}_{M_S}+\psi^{\rho}_{22M_L}(\vrho, \vlab)\chi^{\rho}_{M_S}\bigg)$
~~~~~~\\
& $1^{4}D_{\frac{1}{2}^{+}}$  ~~~~~~& $|56, ^{4}10, 2, 2, \frac{1}{2}^{+}\rangle$
~~~~~~& $\sum\limits_{M_L+M_S=M} \langle 2 M_L; \frac{3}{2} M_S|\frac{1}{2} M\rangle \psi^{S}_{22M_L}(\vrho, \vlab)\chi^{S}_{M_S}$
~~~~~~\\
& $1^{4}D_{\frac{3}{2}^{+}}$  ~~~~~~& $|56, ^{4}10, 2, 2, \frac{3}{2}^{+}\rangle$
~~~~~~& $\sum\limits_{M_L+M_S=M} \langle 2 M_L; \frac{3}{2} M_S|\frac{3}{2} M\rangle \psi^{S}_{22M_L}(\vrho, \vlab)\chi^{S}_{M_S}$
~~~~~~\\
&$1^{4}D_{\frac{5}{2}^{+}}$ ~~~~~~&$|56, ^{4}10, 2, 2, \frac{5}{2}^{+}\rangle$
~~~~~~&$\sum\limits_{M_L+M_S=M} \langle 2 M_L; \frac{3}{2} M_S|\frac{5}{2} M\rangle \psi^{S}_{22M_L}(\vrho, \vlab)\chi^{S}_{M_S}$
~~~~~~\\
&$1^{4}D_{\frac{7}{2}^{+}}$ ~~~~~~&$|56, ^{4}10, 2, 2, \frac{7}{2}^{+}\rangle$
~~~~~~&$\sum\limits_{M_L+M_S=M} \langle 2 M_L; \frac{3}{2} M_S|\frac{7}{2} M\rangle \psi^{S}_{22M_L}(\vrho, \vlab)\chi^{S}_{M_S}$
~~~~~~\\
\hline\hline
\end{tabular}}
\end{center}
\end{table*}

\subsection{States classified in the quark model} \label{Wave functions}

%\subsubsection{Total wave functions}

The $\Omega_{ccc}$ and $\Omega_{bbb}$ spectra should satisfy the requirements of the SU(6)$\times$O(3) symmetry.
The states in the SU(6)$\times$O(3) representation up to the $N=2$ shell are given in Table~\ref{wave function table}.
We denote the baryon states as $|N_6, ^{2S+1}N_3, N, L, J^P\rangle$, where $N_6$ stands for the irreducible representation of
spin-flavor SU(6) group, $N_3$ stands for the irreducible representation of flavor SU(3) group,
and $N$, $S$, $L$, and $J^P$ stand for the principal, spin, total orbital angular momentum, and spin-parity
quantum numbers, respectively. The SU(6)$\times$O(3) wave functions, which correspond to the $|N_6, ^{2S+1}N_3, N, L, J^P\rangle$ states,
are also listed in Table~\ref{wave function table}. The $\psi^{\sigma}_{NLM_L}(\vrho, \vlab)$ and $\chi^{\sigma}_{M_S}$ are the spatial and spin wave functions, respectively, where $\sigma (= s, \rho, \lambda, a)$ denotes the representation of the $S_3$ group.
In the spatial wave functions, $\vrho$ and $\vlab$ are the internal Jacobi coordinates.
The explicit forms of the $\psi^{\sigma}_{NLM_L}(\vrho, \vlab)$ and $\chi^{\sigma}_{M_S}$ have been given in the Ref.~\cite{LZ, Xiao:2013xi}.

\begin{figure*}[htbp]
\begin{center}
\centering  \epsfxsize=16.8 cm \epsfbox{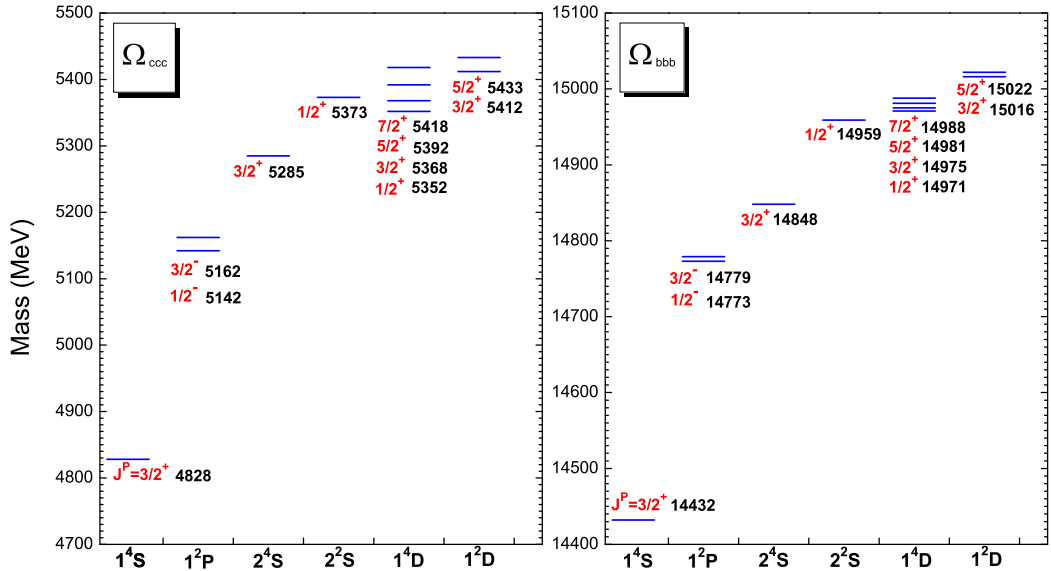}
\vspace{-0.4 cm}\caption{Mass spectra of $\Omega_{ccc}$ and $\Omega_{bbb}$ baryons predicted in this work.} \label{spectra}
\end{center}
\end{figure*}

\subsection{Numerical calculation} \label{Wave functions}

%\subsubsection{Trial spatial wave functions}

The key problem of our numerical calculations is how to deal with the
spatial wave functions. To work out the spatial wave functions,
in this work we expand them in terms of Gaussian basis functions.
The spatial wave function $\psi^{\sigma}_{NLM_L}(\vrho, \vlab)$ may be expressed as~\cite{LZ}
\begin{eqnarray}
\psi^{\sigma}_{NLM_L}(\vrho, \vlab)=\sum_{\begin{subarray}{1}
N=2(n_{\rho}+n_{\lambda})\\
~~~~~~+l_{\rho}+l_{\lambda}\\
M_L=m_\rho+m_\lambda\end{subarray}}C^{n_\rho l_\rho m_\rho }_{n_\lambda l_\lambda m_\lambda}\left[\psi_{n_\rho l_\rho m_\rho}(\vrho) \psi_{n_\lambda l_\lambda m_\lambda}(\vlab)\right]^{\sigma}_{NLM_L}.
\end{eqnarray}
The coefficients $C^{n_\rho l_\rho m_\rho }_{n_\lambda l_\lambda m_\lambda}$ in the spatial wave function $\psi^{\sigma}_{NLM_L}(\vrho, \vlab)$ up to the $N=2$ shell have been given in our previous work~\cite{LZ}. In the above equation, $\psi_{n_\rho l_\rho m_\rho}(\vrho)$ and $\psi_{n_\lambda l_\lambda m_\lambda}(\vlab)$ stand for the spatial wave functions of the $\rho$- and $\lambda$-mode excitations, respectively.

The radial wave functions of the $\rho$- and $\lambda$-mode excitations, $R_{n_\xi l_\xi}(\mathbf{\xi})$ ($\xi=\rho,\lambda$), are expanded by a series of Gaussian basis functions~\cite{LZ}:
\begin{equation}\label{spatial function1}
R_{n_\xi l_\xi}(\xi)=\sum_{\ell=1}^n\mathcal{C}_{\xi\ell}~\phi_{n_\xi l_\xi}(d_{\xi\ell},\xi),
\end{equation}
where
\begin{eqnarray}\label{spatial function2}
%\begin{split}
\phi_{n_\xi l_\xi}(d_{\xi\ell},\xi)&=\left(\frac{1}{d_{\xi\ell}}\right)^{\frac{3}{2}}\Bigg[\frac{2^{l_\xi+2-n_\xi}
(2l_\xi+2n_\xi+1)!!}{\sqrt{\pi}n_\xi![(2l_\xi+1)!!]^2}\Bigg]^{\frac{1}{2}}\left(\frac{\xi}{d_{\xi\ell}}\right)^{l_\xi}\nonumber\\
&\times e^{-\frac{1}{2}\left(\frac{\xi}{d_{\xi\ell}}\right)^2}F\left(-n_\xi,l_\xi+\frac{3}{2},\left(\frac{\xi}{d_{\xi\ell}}\right)^2\right).
%\end{split}
\end{eqnarray}
The $F\left(-n_\xi,l_\xi+\frac{3}{2},\left(\frac{\xi}{d_{\xi\ell}}\right)^2\right)$ is the confluent hypergeometric function. The parameter $d_{\xi\ell}$ can be related to the harmonic oscillator frequency $\omega_{\xi\ell}$ with $1/d^2_{\xi\ell}=M_\xi\omega_{\xi\ell}$.
The reduced masses $M_{\rho,\lambda}$ are defined by $M_{\rho}\equiv \frac{2m_1m_2}{(m_1+m_2)}$, $M_{\lambda}\equiv\frac{3(m_1+m_2)m_3}{2(m_1+m_2+m_3)}$. On the other hand, the harmonic oscillator frequency $\omega_{\xi\ell}$ can be related to the harmonic oscillator stiffness factor $K_{\ell}$ with $\omega_{\xi\ell}=\sqrt{3K_\ell/M_\xi}$~\cite{Xiao:2013xi}. For the identical quark system, one has $d_{\rho\ell}=d_{\lambda\ell}=d_{\ell}=(3m_QK_\ell)^{-1/4}$, where $m_Q$ stands for the constituent mass of the charm or bottom quark.
With this relation, the spatial wave function $\psi^{\sigma}_{NLM_L}(\vrho, \vlab)$ can be simply expanded as
\begin{eqnarray}\label{spatial function3}
\psi^{\sigma}_{NLM_L}(\vrho, \vlab)=\sum_\ell \mathcal{C}_{\ell}\psi^{\sigma}_{NLM_L}(d_\ell,\vrho, \vlab),
\end{eqnarray}
where $\psi^{\sigma}_{NLM_L}(d_\ell,\vrho, \vlab)$ stands for the trial harmonic oscillator functions,
\begin{eqnarray}\label{spatial function3}
\psi^{\sigma}_{NLM_L}(d_\ell,\vrho, \vlab)=\sum_{\begin{subarray}{1}
N=2n_{\rho}+2n_{\lambda}+l_{\rho}+l_{\lambda}\\
M_L=m_\rho+m_\lambda\end{subarray}}C^{n_\rho l_\rho m_\rho }_{n_\lambda l_\lambda m_\lambda}~~~~~~~~~~~~~~~~~~\nonumber\\
\left[\phi_{n_\rho l_\rho}(d_{\ell},\rho)\phi_{n_\lambda l_\lambda}(d_{\ell},\lambda)Y_{l_\rho m_\rho}(\hat{\rho})Y_{l_\lambda m_\lambda}(\hat{\lambda})\right]^\sigma_{NLM_L}.
\end{eqnarray}

%\subsection{Results} \label{mass results}

To solve the Schr\"{o}dinger equation, the variation principle is adopted in this work.
Following the method used in Refs.~\cite{Liu:2019zuc, Hiyama:2003cu}, the oscillator length $d_\ell$ is set to be
\begin{equation}\label{geometric progression}
d_\ell=d_1a^{\ell-1}\ \ \ (\ell=1,...,n),
\end{equation}
where $n$ is the number of Gaussian functions, and $a$ is the ratio coefficient. There are three parameters $\{d_1,d_n,n\}$ to be determined through variation method. It is found that when we take parameters \{0.068fm, 2.711fm, 15\} and \{0.050fm, 2.016fm, 15\} for $\Omega_{ccc}$ baryons and $\Omega_{bbb}$ baryons, respectively, we will obtain stable solutions for the $\Omega_{ccc}$ and $\Omega_{bbb}$ baryons.

Finally, the problem of solving the Schr\"{o}dinger equation become
a problem of solving the generalized matrix eigenvalues of the following equation
\begin{equation}\label{eigenvalue problem}
\sum_{\ell=1}^{n}\sum_{\ell'=1}^{n}(H_{\ell\ell'}-E_\ell N_{\ell\ell'})\mathcal{C}_{\ell'}^\ell=0,
\end{equation}
where $H_{\ell\ell'}\equiv\left\langle \Psi(d_\ell')\Big|H\Big|\Psi(d_\ell)\right\rangle$ and $N_{\ell\ell'}\equiv \left\langle\Psi(d_\ell')\Big|\Psi(d_\ell)\right\rangle$. The function $\Psi(d_\ell)$ is given by
\begin{equation}\label{spatial-spin2}
\Psi(d_\ell)=\sum_{M_L+M_S=M}\langle LM_L,SM_S|JM\rangle\psi^{\sigma}_{NLM_L}(d_\ell,\vrho, \vlab)\chi^{\sigma}_{M_S}.
\end{equation}
The calculations of matrix elements $H_{\ell\ell'}$ and $N_{\ell\ell'}$ have been detailed discussed in Ref.~\cite{LZ}.
The physical state corresponds to the solution with a minimum energy $E_{m}$. By solving this generalized matrix eigenvalue problem, the masses and spacial wave functions of the $\Omega_{ccc}$ and $\Omega_{bbb}$ baryons can be determined.

The predicted masses of the $\Omega_{ccc}$ and $\Omega_{bbb}$ baryons up to $N=2$ shell have been given in Table~\ref{mass table} and also shown in Fig~\ref{Quark model parameters}.

\begin{table}[htp]
\begin{center}
\caption{\label{mass table} Masses (MeV) of the $\Omega_{ccc}$ and $\Omega_{bbb}$ baryons predicted in present work. }
\scalebox{1.0}{
\begin{tabular}{cccccccccccccccccccccc}\hline\hline
~~~~&$n^{2S+1}L_{J^P}$              ~~~~&$|N_6, ^{2S+1}N_3, N, L, J^P\rangle$            ~~~~~~~~~~~~&$\Omega_{ccc}$      ~~~~~~~~~~~~&$\Omega_{bbb}$~~~~\\
\hline
~~~~&$1^{4}S_{\frac{3}{2}^{+}}$     ~~~~&$|56, ^{4}10, 0, 0, \frac{3}{2}^{+}\rangle$     ~~~~~~~~~~~~&4828                ~~~~~~~~~~~~&14432~~~~         \\
~~~~&$1^{2}P_{\frac{1}{2}^{-}}$     ~~~~&$|70, ^{2}10, 1, 1, \frac{1}{2}^{-}\rangle$     ~~~~~~~~~~~~&5142                ~~~~~~~~~~~~&14773~~~~         \\
~~~~&$1^{2}P_{\frac{3}{2}^{-}}$     ~~~~&$|70, ^{2}10, 1, 1, \frac{3}{2}^{-}\rangle$     ~~~~~~~~~~~~&5162                ~~~~~~~~~~~~&14779~~~~         \\
~~~~&$2^{2}S_{\frac{1}{2}^{+}}$     ~~~~&$|70, ^{2}10, 2, 0, \frac{1}{2}^{+}\rangle$     ~~~~~~~~~~~~&5373                ~~~~~~~~~~~~&14959~~~~         \\
~~~~&$2^{4}S_{\frac{3}{2}^{+}}$     ~~~~&$|56, ^{4}10, 2, 0, \frac{3}{2}^{+}\rangle$     ~~~~~~~~~~~~&5285                ~~~~~~~~~~~~&14848~~~~         \\
~~~~&$1^{2}D_{\frac{3}{2}^{+}}$     ~~~~&$|70, ^{2}10, 2, 2, \frac{3}{2}^{+}\rangle$     ~~~~~~~~~~~~&5412                ~~~~~~~~~~~~&15016~~~~         \\
~~~~&$1^{2}D_{\frac{5}{2}^{+}}$     ~~~~&$|70, ^{2}10, 2, 2, \frac{5}{2}^{+}\rangle$     ~~~~~~~~~~~~&5433                ~~~~~~~~~~~~&15022~~~~         \\
~~~~&$1^{4}D_{\frac{1}{2}^{+}}$     ~~~~&$|56, ^{4}10, 2, 2, \frac{1}{2}^{+}\rangle$     ~~~~~~~~~~~~&5352                ~~~~~~~~~~~~&14971~~~~         \\
~~~~&$1^{4}D_{\frac{3}{2}^{+}}$     ~~~~&$|56, ^{4}10, 2, 2, \frac{3}{2}^{+}\rangle$     ~~~~~~~~~~~~&5368                ~~~~~~~~~~~~&14975~~~~         \\
~~~~&$1^{4}D_{\frac{5}{2}^{+}}$     ~~~~&$|56, ^{4}10, 2, 2, \frac{5}{2}^{+}\rangle$     ~~~~~~~~~~~~&5392                ~~~~~~~~~~~~&14981~~~~         \\
~~~~&$1^{4}D_{\frac{7}{2}^{+}}$     ~~~~&$|56, ^{4}10, 2, 2, \frac{7}{2}^{+}\rangle$     ~~~~~~~~~~~~&5418                ~~~~~~~~~~~~&14988~~~~         \\
\hline\hline
\end{tabular}}
\end{center}
\end{table}

\section{radiative decays}\label{radiative decay}

%\subsection{Framework} \label{radiative decay in CQM}

In this work the radiative decays of the $\Omega_{ccc}$ and $\Omega_{bbb}$ baryon states are evaluated within a nonrelativistic constituent
quark model developed in our previous study of the heavy quarkonia~\cite{Deng:2016stx,Deng:2016ktl}.
This model has been extended to deal with the radiative decays of the $B_c$ meson states~\cite{Li:2019tbn}, $\Omega$ baryon states~\cite{LZ}, singly baryon states~\cite{Yao:2018jmc,Wang:2017kfr,Wang:2017hej}.

In this model, the quark-photon electromagnetic (EM) coupling at the tree level is adopted as
\begin{eqnarray}\label{he}
H_e=-\sum_j
e_{j}\bar{\psi}_j\gamma^{j}_{\mu}A^{\mu}(\mathbf{k},\mathbf{r}_j)\psi_j,
\end{eqnarray}
where $A^{\mu}$ is the photon field with three momentum $\mathbf{k}$, while
$\mathbf{r}_j$ and $e_j$ stand for the coordinate and charge of the
$j$th quark field $\psi_j$.

In order to match the nonrelativistic wave functions of the baryons, we should adopt the nonrelativistic form of Eq.~(\ref{he}) in the calculations. Including the effects of the binding potential between quarks~\cite{Brodsky:1968ea}, the nonrelativistic expansion of $H_e$ may be written as ~\cite{Li:1994cy,Li:1997gd,Close:1970kt}
\begin{equation}\label{he2}
h_{e}\simeq\sum_{j}\left[e_{j}\mathbf{r}_{j}\cdot\veps-\frac{e_{j}}{2m_{j}
}\vsig_{j}\cdot(\veps\times\hat{\mathbf{k}})\right]e^{-i\mathbf{k}\cdot\mathbf{r}_j},
\end{equation}
where $m_j$ and $\vsig_j$ stand for the constituent mass and Pauli spin vector for the $j$th quark. The vector $\veps$ is the polarization vector of the photon. This nonrelativistic EM transition operator has between widely applied to meson photoproduction reactions~\cite{Li:1994cy,Li:1995si,Li:1997gd,
Zhao:2001kk, Saghai:2001yd, Zhao:2002id, He:2008ty, He:2008uf,
He:2010ii, Zhong:2011ti, Zhong:2011ht, Xiao:2015gra}.

Then, the standard helicity transition amplitude $\mathcal{A}_{\lambda}$ between the initial baryon state $|\mathcal{B}\rangle$ and the final baryon state $|\mathcal{B}\rangle$ can be calculated by
\begin{eqnarray}\label{amp3}
\mathcal{A}_{\lambda}&=&-i\sqrt{\frac{\omega_\gamma}{2}}\langle \mathcal{B}' | h_{e}|\mathcal{B}
\rangle.
\end{eqnarray}
where $\omega_{\gamma}$ is the photon energy.

Finally, we can calculate the EM decay width by
\begin{equation}\label{dww}
\Gamma_{\gamma}=\frac{|\mathbf{k}|^2}{\pi}\frac{2}{2J_i+1}\frac{M_{f}}{M_{i}}\sum_{J_{fz},J_{iz}}|\mathcal{A}_{J_{fz},J_{iz}}|^2,
\end{equation}
where $J_i$ is the total angular momentum of an initial meson,
$J_{fz}$ and $J_{iz}$ are the components of the total angular
momenta along the $z$ axis of initial and final mesons,respectively.

%\subsection{Results} \label{radiative decay results}

In our calculations, the masses and the wave functions of the $\Omega_{ccc}$ and $\Omega_{bbb}$ baryon states are adopted by solving the Schr\"{o}dinger equation in Sec.II. The radiative decay widths of $\Omega_{ccc}$ and $\Omega_{bbb}$ baryons up to $N=2$ are listed in Table~\ref{the radiative decay widths}. For simplicity one can fit the numerical wave functions with a single Gaussian (SG) form by reproducing the root-mean-square radius of the $\rho$-mode excitations. The determined harmonic oscillator strength parameters, $\alpha$, for the $\Omega_{ccc}$ and $\Omega_{bbb}$ baryon states are listed in Table~\ref{alpha and rms}. With the the SG effective wave functions, we also calculated the radiative decay widths of the $\Omega_{ccc}$ and $\Omega_{bbb}$ baryon states, these results are listed in Table~\ref{the radiative decay widths} for a comparison.
From Table~\ref{the radiative decay widths}, it is find that the partial widths obtained with the SG effective wave functions show less differences  with those obtained with the real numerical wave functions.

\begin{table*}[htp]
\begin{center}
\caption{\label{the radiative decay widths} Partial widths (keV) of radiative decays for the $\Omega_{ccc}$ and $\Omega_{bbb}$ baryons up to $N=2$ shell. Case I and Case II stand the results obtained within the real numerical wave functions and the single Gaussian wave functions, respectively.}
\scalebox{1.0}{
\begin{tabular}{ccccccccccccccccccc}\hline\hline
~~~~~&                                 ~~~~~~&\multicolumn{2}{c}{$\underline{~~\Gamma[\Omega_{ccc}(1^{4}S_{3/2^+})\gamma]~~}$}     ~~~~~~&&
~~~~~~~~~~&                            ~~~~~~&\multicolumn{2}{c}{$\underline{~~\Gamma[\Omega_{ccc}(1^{4}S_{3/2^+})\gamma]~~}$}     ~~~~~~&&~~~~~\\
~~~~~&Initial state                    ~~~~~~&Case I & Case II                                                                              ~~~~~~&&
~~~~~~~~~~&Initial state               ~~~~~~&Case I & Case II                                                                               ~~~~~~&&~~~~~\\
\hline
~~~~~&$\Omega_{ccc}(1^{2}P_{1/2^-})$   ~~~~~~&3.10     &2.78       ~~~~~~&       &          ~~~~~~~~~~&$\Omega_{bbb}(1^{2}P_{1/2^-})$   ~~~~~&0.035&0.028            ~~~~~~&         &                                 ~~~~~\\
~~~~~&$\Omega_{ccc}(1^{2}P_{3/2^-})$   ~~~~~~&4.07     &3.64       ~~~~~~&       &          ~~~~~~~~~~&$\Omega_{bbb}(1^{2}P_{3/2^-})$   ~~~~~&0.038&0.031            ~~~~~~&         &                                 ~~~~~\\
\hline\hline
~~~~~&                                ~~~~~~&\multicolumn{2}{c}{$\underline{~~\Gamma[\Omega_{ccc}(1^{2}P_{1/2^-})\gamma]~~}$}      ~~~~~~&\multicolumn{2}{c}{$\underline{~~\Gamma[\Omega_{ccc}(1^{2}P_{3/2^-})\gamma]~~}$}
~~~~~~~~~~&                           ~~~~~~&\multicolumn{2}{c}{$\underline{~~\Gamma[\Omega_{bbb}(1^{2}P_{1/2^-})\gamma]~~}$}      ~~~~~~&\multicolumn{2}{c}{$\underline{~~\Gamma[\Omega_{bbb}(1^{2}P_{3/2^-})\gamma]~~}$}~~~~~\\
~~~~~&Initial state                   ~~~~~~&Case I & Case II                                                                               ~~~~~~&Case I & Case II
~~~~~~~~~~&Initial state              ~~~~~~&Case I & Case II                                                                               ~~~~~~&Case I & Case II ~~~~~\\
\hline
~~~~~&$\Omega_{ccc}(2^{2}S_{1/2^+})$  ~~~~~~&20.14    &18.07      ~~~~~~~&27.43  &24.70     ~~~~~~~~~~&$\Omega_{bbb}(2^{2}S_{1/2^+})$   ~~~~~&0.99     &0.82         ~~~~~~&1.46     &1.25                             ~~~~~\\
~~~~~&$\Omega_{ccc}(2^{4}S_{3/2^+})$  ~~~~~~&0.002    &0.001      ~~~~~~~&0.010  &0.009     ~~~~~~~~~~&$\Omega_{bbb}(2^{4}S_{3/2^+})$   ~~~~~&$<0.001$ &$<0.001$     ~~~~~~&$<0.001$ &$<0.001$                         ~~~~~\\
~~~~~&$\Omega_{ccc}(1^{2}D_{3/2^+})$  ~~~~~~&106.88   &98.80      ~~~~~~~&33.58  &31.26     ~~~~~~~~~~&$\Omega_{bbb}(1^{2}D_{3/2^+})$   ~~~~~&7.84     &6.94         ~~~~~~&2.75     &2.44                             ~~~~~\\
~~~~~&$\Omega_{ccc}(1^{2}D_{5/2^+})$  ~~~~~~&0.25     &0.22       ~~~~~~~&122.10 &113.46    ~~~~~~~~~~&$\Omega_{bbb}(1^{2}D_{5/2^+})$   ~~~~~&0.001    &0.001        ~~~~~~&8.42     &7.44                             ~~~~~\\
~~~~~&$\Omega_{ccc}(1^{4}D_{1/2^+})$  ~~~~~~&$<0.001$ &$<0.001$   ~~~~~~~&0.04   &0.04      ~~~~~~~~~~&$\Omega_{bbb}(1^{4}D_{1/2^+})$   ~~~~~&$<0.001$ &$<0.001$     ~~~~~~&$<0.001$ &$<0.001$                         ~~~~~\\
~~~~~&$\Omega_{ccc}(1^{4}D_{3/2^+})$  ~~~~~~&0.38     &0.35       ~~~~~~~&0.02   &0.02      ~~~~~~~~~~&$\Omega_{bbb}(1^{4}D_{3/2^+})$   ~~~~~&0.002    &0.002        ~~~~~~&$<0.001$ &$<0.001$                         ~~~~~\\
~~~~~&$\Omega_{ccc}(1^{4}D_{5/2^+})$  ~~~~~~&0.22     &0.21       ~~~~~~~&0.39   &0.34      ~~~~~~~~~~&$\Omega_{bbb}(1^{4}D_{5/2^+})$   ~~~~~&$<0.001$ &$<0.001$     ~~~~~~&0.002    &0.002                            ~~~~~\\
~~~~~&$\Omega_{ccc}(1^{4}D_{7/2^+})$  ~~~~~~&$<0.001$ &$<0.001$   ~~~~~~~&0.80   &0.74      ~~~~~~~~~~&$\Omega_{bbb}(1^{4}D_{7/2^+})$   ~~~~~&$<0.001$ &$<0.001$     ~~~~~~&0.003    &0.002                            ~~~~~\\
\hline\hline
\end{tabular}}
\end{center}
\end{table*}

\begin{table}[htp]
\begin{center}
\caption{\label{alpha and rms} The effective harmonic oscillator strength parameters $\alpha$ and root-mean-square $\langle r_{12}^2 \rangle^{1/2}$ of $\Omega_{ccc}$ and $\Omega_{bbb}$ baryons up to $N=2$ shell.}
\scalebox{1.0}{
\begin{tabular}{ccccccccccccccccccccccccccccccccccccccccccccccccccccccccc}\hline\hline
~&                                ~&\multicolumn{2}{c}{$\underline{~~~~~~~~~~~~~~~~\Omega_{ccc}~~~~~~~~~~~~~~~~}$}      ~~~~~~&\multicolumn{2}{c}{$\underline{~~~~~~~~~~~~~~~~\Omega_{bbb}~~~~~~~~~~~~~~~~}$}\\
~&$n^{2S+1}L_{J^P}$               ~&$\alpha$(MeV)  ~&$\langle r_{12}^2 \rangle^{1/2}$(fm)                               ~~~~~~&$\alpha$(MeV)  ~&$\langle r_{12}^2 \rangle^{1/2}$(fm)                         \\
\hline
~&$1^{4}S_{\frac{3}{2}^{+}}$      ~&716            ~&0.48                                                               ~~~~~~&1358           ~&0.25                                                         \\
~&$1^{2}P_{\frac{1}{2}^{-}}$      ~&637            ~&0.62                                                               ~~~~~~&1090           ~&0.36                                                         \\
~&$1^{2}P_{\frac{1}{3}^{-}}$      ~&627            ~&0.63                                                               ~~~~~~&1084           ~&0.36                                                         \\
~&$2^{2}S_{\frac{1}{2}^{+}}$      ~&581            ~&0.76                                                               ~~~~~~&968            ~&0.46                                                         \\
~&$2^{4}S_{\frac{3}{2}^{+}}$      ~&581            ~&0.76                                                               ~~~~~~&992            ~&0.44                                                         \\
~&$1^{2}D_{\frac{3}{2}^{+}}$      ~&584            ~&0.76                                                               ~~~~~~&955            ~&0.46                                                         \\
~&$1^{2}D_{\frac{5}{2}^{+}}$      ~&575            ~&0.77                                                               ~~~~~~&947            ~&0.47                                                         \\
~&$1^{4}D_{\frac{1}{2}^{+}}$      ~&597            ~&0.74                                                               ~~~~~~&973            ~&0.45                                                         \\
~&$1^{4}D_{\frac{3}{2}^{+}}$      ~&589            ~&0.75                                                               ~~~~~~&968            ~&0.46                                                         \\
~&$1^{4}D_{\frac{5}{2}^{+}}$      ~&579            ~&0.76                                                               ~~~~~~&963            ~&0.46                                                         \\
~&$1^{4}D_{\frac{7}{2}^{+}}$      ~&569            ~&0.78                                                               ~~~~~~&955            ~&0.46                                                         \\
\hline\hline
\end{tabular}}
\end{center}
\end{table}

\section{discussions} \label{discussions}

\subsection{Ground states}

\begin{figure*}[htbp]
\begin{center}
\centering  \epsfxsize=16 cm \epsfbox{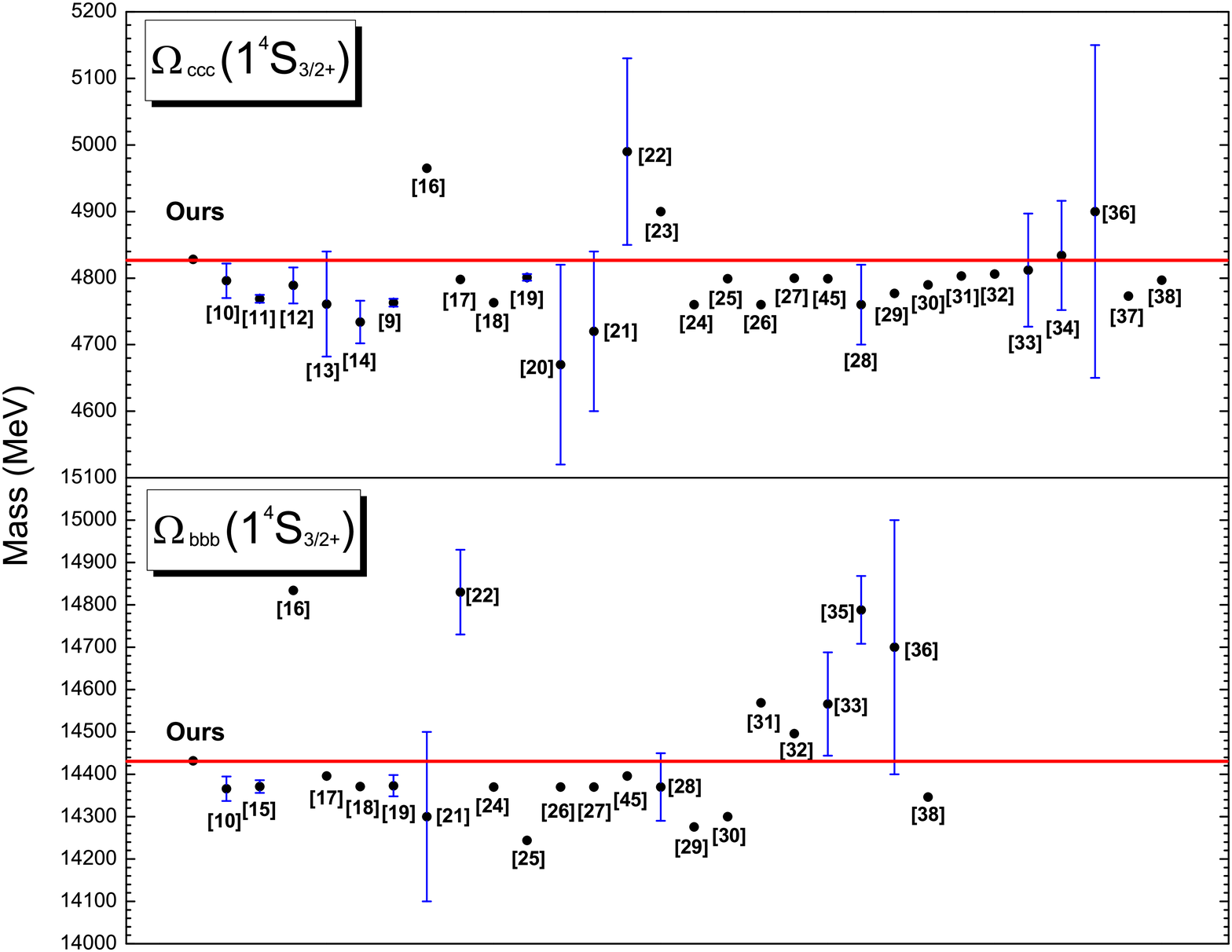}
\vspace{-1.0 cm}\caption{A comparison of the masses of the ground states $\Omega_{ccc}$ and $\Omega_{bbb}$ from various model predictions.} \label{ground}
\end{center}
\end{figure*}

For the ground states $\Omega_{ccc}$ and $\Omega_{bbb}$, our predicted masses are
$\sim 4828$ MeV and $\sim 14432$ MeV, respectively.
There are many predictions of the masses of $\Omega_{ccc}$ and $\Omega_{bbb}$ in the literature~\cite{ref1,ref2,ref3,ref5,ref6,ref7,ref8,ref9,ref10,ref11,ref12,ref13,ref14,ref15,ref16,
ref17,ref18,ref19,ref20,ref21,ref22,ref23,ref24,ref25,ref26,ref27,ref28,ref29,ref30,ref31}.
For a comparison, our results and those of other works are collected in Table~\ref{ground mass table} and also shown in Fig~\ref{ground}.

It is found that in most of the studies the masses of the ground states $\Omega_{ccc}$ and $\Omega_{bbb}$ are predicted to
be in the range of $\sim 4800\pm50$ MeV and $\sim 14410\pm170$ MeV, respectively.
Our predicted masses are reasonably consistent with the previous studies, although our results are slightly larger most
of the other predictions (see Fig~\ref{ground}). Compared with the results of the lattice QCD, it is found
that our predicted mass for $\Omega_{ccc}$ just lies the upper limit of the predictions in Refs.~\cite{ref12,ref14,ref15},
while our predicted mass for $\Omega_{bbb}$ is about 60 MeV above the predictions in Refs.~\cite{ref12,ref18}.

\begin{table}[htp]
\begin{center}
\caption{\label{ground mass table} Our predicted masses (MeV) of the ground states $\Omega_{ccc}$ and $\Omega_{bbb}$ compared with
those of other works. }
\scalebox{1.0}{
\begin{tabular}{ccccccccccccccccccccccccccccccccc}\hline\hline
~&Method                                         ~&$\Omega_{ccc}(1^4S_{3/2^+})$            ~&$\Omega_{bbb}(1^4S_{3/2^+})$~  \\
\hline
~&NRCQM                     ~(Ours)              ~&4828                                    ~&14432~                         \\
~&Lattice QCD               ~\cite{ref12}        ~&4796(8)(18)                             ~&14366(9)(20)~                  \\
~&Lattice QCD               ~\cite{ref13}        ~&4769(6)                                 ~&$\cdots$~                      \\
~&Lattice QCD               ~\cite{ref14}        ~&4789(6)(21)                             ~&$\cdots$~                      \\
~&Lattice QCD               ~\cite{ref15}        ~&4761(52)(21)(6)                         ~&$\cdots$~                      \\
~&Lattice QCD               ~\cite{ref16}        ~&4734(12)(11)(9)                         ~&$\cdots$~                      \\
~&Lattice QCD               ~\cite{ref17}        ~&4763(6)                                 ~&$\cdots$~                      \\
~&Lattice QCD               ~\cite{ref18}        ~&$\cdots$                                ~&$14371\pm4\pm11$~              \\
~&NRCQM                     ~\cite{ref2}         ~&4965                                    ~&14834~                         \\
~&NRCQM                     ~\cite{ref21}        ~&4798                                    ~&14396~                         \\
~&NRCQM                     ~\cite{ref23}        ~&4763                                    ~&14371~                         \\
~&NRCQM                     ~\cite{ref25}        ~&$4801\pm5$                              ~&$14373\pm25$~                  \\
~&QCD Sum Rule              ~\cite{ref7}         ~&$4670\pm150$                            ~&$13280\pm100$~                 \\
~&QCD Sum Rule              ~\cite{ref10}        ~&$4720\pm120$                            ~&$14300\pm200$~                 \\
~&QCD Sum Rule              ~\cite{ref27}        ~&$4990\pm140$                            ~&$14830\pm100$~                 \\
~&Faddeev Equation          ~\cite{ref19}        ~&4900                                    ~&13800~                         \\
~&Faddeev Equation          ~\cite{ref29}        ~&4760                                    ~&14370~                         \\
~&Faddeev Equation          ~\cite{ref31}        ~&4799                                    ~&14244~                         \\
~&Diquark Model             ~\cite{ref9}         ~&4760                                    ~&14370~                         \\
~&Diquark Model             ~\cite{ref30}        ~&4800                                    ~&14370~                         \\
~&Variational Method        ~\cite{ref4}         ~&4799                                    ~&14396~                         \\
~&Variational Method        ~\cite{ref28}        ~&$4760\pm60$                             ~&$14370\pm80$~                  \\
~&Bag model                 ~\cite{ref3}         ~&4777                                    ~&14276~                         \\
~&Bag model                 ~\cite{ref8}         ~&4790                                    ~&14300~                         \\
~&RQM                       ~\cite{ref11}        ~&4803                                    ~&14569~                         \\
~&HCQM                      ~\cite{ref1}         ~&4806                                    ~&14496~                         \\
~&HCQM                      ~\cite{ref22}        ~&$4812\pm85$                             ~&$14566\pm122$~                 \\
~&Regge Theory              ~\cite{ref5}         ~&$4834^{+82}_{-81}$                      ~&$\cdots$~                      \\
~&Regge Theory              ~\cite{ref6}         ~&$\cdots$                                ~&$14788\pm80$~                  \\
~&NRQCD                     ~\cite{ref20}        ~&4900(250)                               ~&14700(300)~                    \\
~&Bathe-Salpeter Equation   ~\cite{ref24}        ~&4773                                    ~&$\cdots$~                      \\
~&RGPEP                     ~\cite{ref26}        ~&4797                                    ~&14346~                         \\
\hline\hline
\end{tabular}}
\end{center}
\end{table}

\begin{table*}[htp]
\begin{center}
\caption{\label{1P mass table} Our predicted masses (MeV) of the 1$P$-wave $\Omega_{ccc}$ and $\Omega_{bbb}$ states compared with
those of other works. }
\scalebox{1.0}{
\begin{tabular}{ccccccccccccccccccccccccccccccccc}\hline\hline
~~~~&Method                   ~~~~~~~~&$\Omega_{ccc}(1^2P_{1/2^-})$  ~~~~~~~~&$\Omega_{ccc}(1^2P_{3/2^-})$  ~~~~~~~~&$\Omega_{bbb}(1^2P_{1/2^-})$  ~~~~~~~~&$\Omega_{bbb}(1^2P_{3/2^-})$~~~~  \\
\hline
~~~~&NRCQM                     ~(Ours)          ~~~~~~~~&5142                          ~~~~~~~~&5162                          ~~~~~~~~&14773                         ~~~~~~~~&14779~~~~                         \\
~~~~&Lattice QCD               ~\cite{ref17}    ~~~~~~~~&5120(9)                       ~~~~~~~~&5124(13)                      ~~~~~~~~&$\cdots$                      ~~~~~~~~&$\cdots$~~~~                      \\
~~~~&Lattice QCD               ~\cite{ref18}    ~~~~~~~~&$\cdots$                      ~~~~~~~~&$\cdots$                      ~~~~~~~~&$14706.3\pm9.8\pm18.4$        ~~~~~~~~&$14714\pm9.5\pm18.2$~~~~          \\
~~~~&NRCQM                     ~\cite{ref2}     ~~~~~~~~&5155                          ~~~~~~~~&5160                          ~~~~~~~~&14975                         ~~~~~~~~&14976~~~~                         \\
~~~~&NRCQM                     ~\cite{ref21}    ~~~~~~~~&5129                          ~~~~~~~~&5129                          ~~~~~~~~&14688                         ~~~~~~~~&14688~~~~                         \\
~~~~&QCD Sum Rule              ~\cite{ref10}    ~~~~~~~~&$\cdots$                      ~~~~~~~~&$4900\pm100$                  ~~~~~~~~&$\cdots$                      ~~~~~~~~&$14900\pm200$~~~~                 \\
~~~~&QCD Sum Rule              ~\cite{ref27}    ~~~~~~~~&$\cdots$                      ~~~~~~~~&$5110\pm150$                  ~~~~~~~~&$\cdots$                      ~~~~~~~~&$14950\pm110$~~~~                 \\
~~~~&Faddeev Equation          ~\cite{ref29}    ~~~~~~~~&$\cdots$                      ~~~~~~~~&5027                          ~~~~~~~~&$\cdots$                      ~~~~~~~~&14771~~~~                         \\
~~~~&Bag model                 ~\cite{ref8}     ~~~~~~~~&5140                          ~~~~~~~~&$\cdots$                      ~~~~~~~~&14660                         ~~~~~~~~&$\cdots$ ~~~~                     \\
~~~~&HCQM                      ~\cite{ref1}     ~~~~~~~~&5002                          ~~~~~~~~&4982                          ~~~~~~~~&14941                         ~~~~~~~~&14935~~~~                         \\
~~~~&Regge theory              ~\cite{ref5}     ~~~~~~~~&$\cdots$                      ~~~~~~~~&$5073^{+109}_{-107}$          ~~~~~~~~&$\cdots$                      ~~~~~~~~&$\cdots$ ~~~~                     \\
~~~~&Regge theory              ~\cite{ref6}     ~~~~~~~~&$\cdots$                      ~~~~~~~~&$\cdots$                      ~~~~~~~~&$\cdots$                      ~~~~~~~~&$15055\pm101$~~~~                 \\
~~~~&Bathe-Salpeter Equation   ~\cite{ref24}    ~~~~~~~~&5019                          ~~~~~~~~&5014                          ~~~~~~~~&$\cdots$                      ~~~~~~~~&$\cdots$ ~~~~                     \\
\hline\hline
\end{tabular}}
\end{center}
\end{table*}

\begin{figure*}[htbp]
\begin{center}
\centering  \epsfxsize=16 cm \epsfbox{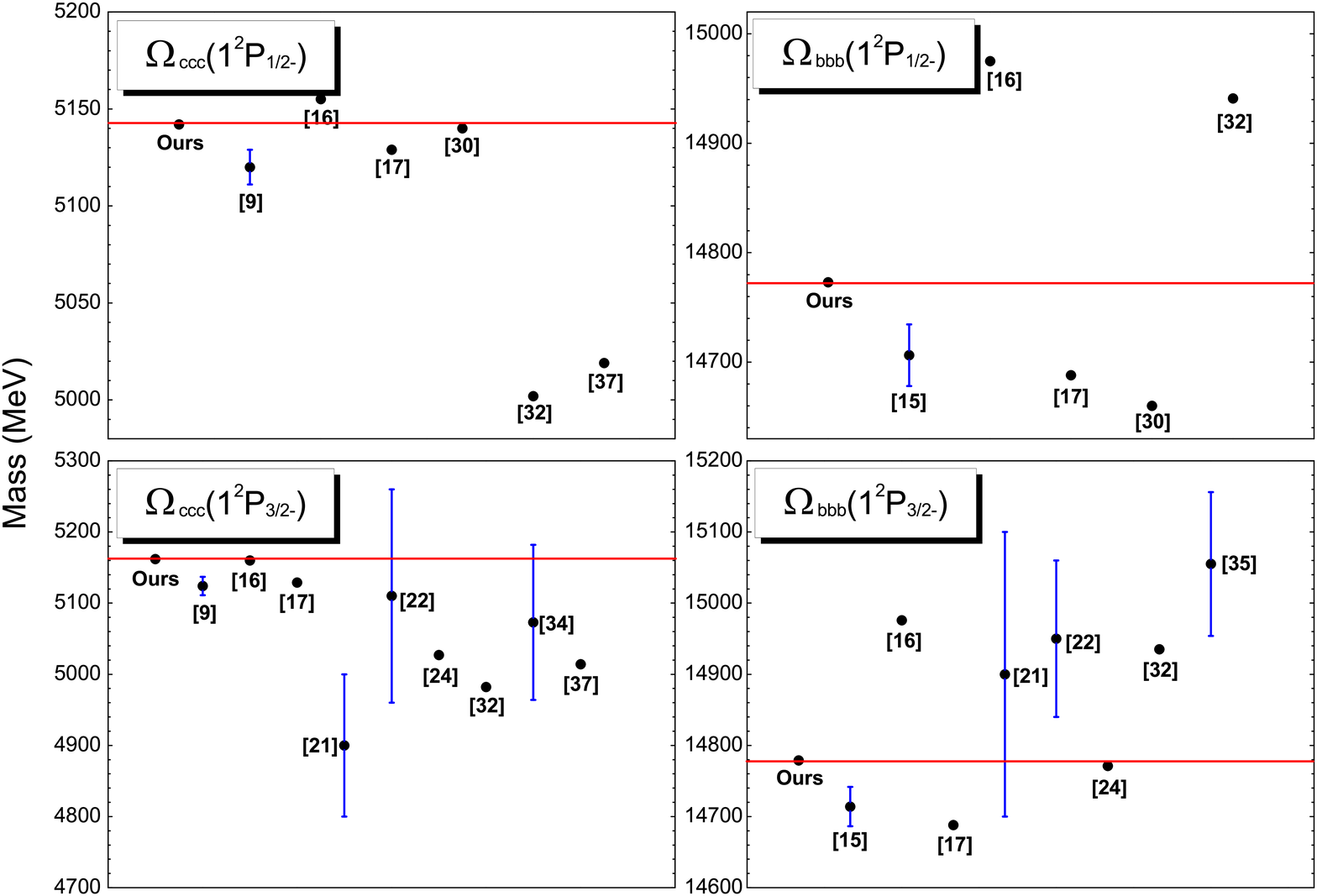}
\vspace{-1.0 cm}\caption{ A comparison of the masses of the $1P$-wave $\Omega_{ccc}$ and $\Omega_{bbb}$ states from various model predictions.} \label{1P}
\end{center}
\end{figure*}

\subsection{$1P$-wave states}

There are two $1P$-wave $\Omega_{QQQ}$ ($Q=c,b$) states with $J^P=1/2^-$ and $J^P=3/2^-$ according to the quark model classification (see Table~\ref{wave function table}). For a comparison, our predicted masses of the 1$P$-wave $\Omega_{ccc}$ and $\Omega_{bbb}$ states together with those of other theoretical predictions have been listed in Table~\ref{1P mass table} and shown in Figure~\ref{1P}.
Our predictions of the radiative decay properties of the $1P$-wave $\Omega_{ccc}$ and $\Omega_{bbb}$ states are also given in Table~\ref{the radiative decay widths}.

\subsubsection{$\Omega_{ccc}(1P)$ states}

In our calculations, the masses of the $1P$-wave states $\Omega_{ccc}(1^2P_{1/2^-})$ and $\Omega_{ccc}(1^2P_{3/2^-})$ are predicted to
be $\sim 5142$ MeV and $\sim 5162$ MeV, respectively, which are close to the values
$\sim 5120(9)$ MeV $\sim 5124(13)$ MeV from the Lattice QCD calculation~\cite{ref17}.
Our results are also compatible with the other model predictions in Refs.~\cite{ref2,ref21,ref8,ref27,ref5}.
The mass splitting between $\Omega_{ccc}(1^2P_{1/2^-})$ and $\Omega_{ccc}(1^2P_{3/2^-})$ might be small.
With a simplified phenomenological spin-orbit potential as adopted in the study of the
$\Omega$ spectrum in our previous work~\cite{LZ}, we predict that the mass splitting between these
two $1P$-wave states might be $\sim 20$ MeV, which is slightly larger than the value of several MeV
predicted in the literature~\cite{ref17,ref2}.

The decays of the $1P$-wave $\Omega_{ccc}(1^2P_{1/2^-})$ and $\Omega_{ccc}(1^2P_{3/2^-})$
states may be dominated by the radiative transitions into the ground $1S$-wave state $\Omega_{ccc}$,
for their OZI-allowed two body strong decay processes are absence. We further estimate the radiative decays of
the $\Omega_{ccc}(1^2P_{1/2^-})$ and $\Omega_{ccc}(1^2P_{3/2^-})$ states by using the wave functions
calculated from the potential model. It is found that both $\Omega_{ccc}(1^2P_{1/2^-})$ and
$\Omega_{ccc}(1^2P_{3/2^-})$ have a comparable radiative decay width into the ground $1S$-wave state
$\Omega_{ccc}$, i.e.,
\begin{eqnarray}
\Gamma[\Omega_{ccc}(1^2P_{1/2^-})\to \Omega_{ccc}\gamma]= 3.10 ~\text{keV},\\
\Gamma[\Omega_{ccc}(1^2P_{3/2^-})\to \Omega_{ccc}\gamma]= 4.07 ~\text{keV}.
\end{eqnarray}
The radiative transitions $\Omega_{ccc}(1^2P_{1/2^-},1^2P_{3/2^-})\to \Omega_{ccc}\gamma$ may
be crucial to established them in future experiments. It should be mentioned
that few studies radiative decay properties of the excited $\Omega_{ccc}$
are found in the literature. More theoretical analysis is need to better understand
these $1P$-wave states.

\subsubsection{$\Omega_{bbb}(1P)$ states}

The masses of the $1P$-wave states $\Omega_{bbb}(1^2P_{1/2^-})$ and $\Omega_{bbb}(1^2P_{3/2^-})$ are predicted to be $\sim 14773$ MeV and
$\sim 14779$ MeV, respectively. Our predictions are about $40-100$ MeV larger than those predictions in Refs.~\cite{ref18,ref21,ref8}, while about 200 MeV smaller than the those predictions in Refs.~\cite{ref2,ref1}. It should be mentioned that our predicted mass of $\Omega_{bbb}(1^2P_{3/2^-})$ is in good agreement with that predicted with Faddeev Equation~\cite{ref29}.
Furthermore, our predicted mass splitting between $\Omega_{bbb}(1^2P_{1/2^-})$ and $\Omega_{bbb}(1^2P_{3/2^-})$, $\sim 6$ MeV, is similar to the Lattice QCD prediction of $\sim 8$ MeV in Ref.~\cite{ref18}.

In the following we give a brief discussion of the relations of the mass splitting between the $1P$-wave states in the $\Omega_{QQQ}$ ($Q\in\{s, c, b\}$) baryon spectrum. If the mass splitting between two $1P$-wave states is due to the spin-orbit interaction, from Eq.~(\ref{voge LS}) one finds that the mass splitting $\Delta m [\Omega_{QQQ}(1P)]\propto \frac{1}{m_Q^2}\left\langle \frac{1}{ \rho^2+\lambda^2 }\right\rangle$, where $m_Q$ is the mass of constituent quark $Q$. With a simple harmonic oscillator wave function, one can relate the element matrix $\left\langle \frac{1}{ \rho^2+\lambda^2 }\right\rangle$ to the harmonic oscillator strength parameter $\alpha$. One further finds that $\left\langle \frac{1}{ \rho^2+\lambda^2 }\right\rangle \propto \alpha^2$. Then we obtain an useful relation for the mass splitting:
\begin{eqnarray}
\Delta m [\Omega_{QQQ}(1P)] \propto \bigg(\frac{\alpha}{m_Q}\bigg)^2.
\end{eqnarray}
Taking the constituent quark masses and effective harmonic oscillator strength parameters $\alpha$ for the $1P$-wave $\Omega$, $\Omega_{ccc}$, and $\Omega_{bbb}$ states determined in present work and our previous work~\cite{LZ}, we obtain the following ratios
\begin{equation}
\Delta m [\Omega(1P)]:\Delta m [\Omega_{ccc}(1P)]:\Delta m [\Omega_{bbb}(1P)] \simeq 9: 3: 1.
\end{equation}
Future experimental measurements of these ratios may provide a crucial test for the spin-orbit interactions adopted in present work.

The radiative decay properties of the $\Omega_{bbb}(1^2P_{1/2^-})$ and $\Omega_{bbb}(1^2P_{3/2^-})$ states
are also estimated in present work by using the wave functions calculated from the potential model.
The partial widths for the $\Omega_{bbb}(1^2P_{1/2^-})$ and $\Omega_{bbb}(1^2P_{3/2^-})$ states decaying into the ground state $\Omega_{bbb}$ are predicted to be
\begin{eqnarray}
\Gamma[\Omega_{bbb}(1^2P_{1/2^-})\to \Omega_{bbb}\gamma]= 0.035 ~\text{keV},\\
\Gamma[\Omega_{bbb}(1^2P_{3/2^-})\to \Omega_{bbb}\gamma]= 0.038 ~\text{keV}.
\end{eqnarray}
Combing the partial widths of the $1P$-wave $\Omega_{ccc}$ states, one find
\begin{eqnarray}
\mathcal{R}=\frac{\Gamma[\Omega_{bbb}(1P)\to \Omega_{bbb}(1S)\gamma]}{\Gamma[\Omega_{ccc}(1P)\to \Omega_{ccc}(1S)\gamma]}\simeq \frac{1}{100}.
\end{eqnarray}
Since the partial widths for the $1P$-wave $\Omega_{ccc}$ states are about two orders of
magnitude smaller than those corresponding processes of the $1P$-wave $\Omega_{bbb}$ states,
the radiative decay process of $\Omega_{bbb}(1P)\to \Omega_{bbb}\gamma$ may be more difficultly observed than $\Omega_{ccc}(1P)\to \Omega_{ccc}\gamma$.

\subsection{$1D$-wave states}

\begin{table*}[htp]
\begin{center}
\caption{\label{2S1D mass table} Our predicted masses (MeV) of the 2$S$- and 1$D$-wave $\Omega_{ccc}$ and $\Omega_{bbb}$ states compared with
those of other works. }
\scalebox{1.0}{
\begin{tabular}{ccccccccccccccccccccccccccccccccccccccccccccccccccccccccc}\hline\hline
~&                                ~&\multicolumn{5}{c}{$\underline{~~~~~~~~~~~~~~~~~~~~~~~~~~~~~~~~~~~~~~~~\Omega_{ccc}~~~~~~~~~~~~~~~~~~~~~~~~~~~~~~~~~~~~~~~~}$}
                             ~~~~~~&\multicolumn{5}{c}{$\underline{~~~~~~~~~~~~~~~~~~~~~~~~~~~~~~~~~~~~~~~~~~~~~~\Omega_{bbb}~~~~~~~~~~~~~~~~~~~~~~~~~~~~~~~~~~~~~~~~~~~~~~}$}\\
~&                                ~&NRCQM  ~&Lattice            ~&NRCQM             ~&NRCQM              ~&HCQM              ~~~~~~&NRCQM  ~&Lattice            ~&NRCQM             ~&NRCQM              ~&HCQM            \\
~&$n^{2S+1}L_{J^P}$               ~&Ours   ~&Ref.~\cite{ref17}  ~&Ref.~\cite{ref2}  ~&Ref.~\cite{ref21}  ~&Ref.~\cite{ref1}  ~~~~~~&Ours   ~&Ref.~\cite{ref18}  ~&Ref.~\cite{ref2}  ~&Ref.~\cite{ref21}  ~&Ref.~\cite{ref1}\\
\hline
~&$2^{2}S_{\frac{1}{2}^{+}}$      ~&5373   ~&5405(14)           ~&5332              ~&$\cdots$           ~&5300              ~~~~~~&14959  ~&$14938\pm18\pm23$  ~&15097             ~&$\cdots$           ~&15163           \\
~&$2^{4}S_{\frac{3}{2}^{+}}$      ~&5285   ~&5317(31)           ~&5313              ~&5286               ~&5300              ~~~~~~&14848  ~&$14840\pm15\pm20$  ~&15089             ~&14805              ~&15163           \\
~&$1^{2}D_{\frac{3}{2}^{+}}$      ~&5412   ~&5465(13)           ~&$\cdots$          ~&$\cdots$           ~&5436              ~~~~~~&15016  ~&$15005\pm18\pm24$  ~&$\cdots$          ~&$\cdots$           ~&15298           \\
~&$1^{2}D_{\frac{5}{2}^{+}}$      ~&5433   ~&5464(15)           ~&5343              ~&$\cdots$           ~&5404              ~~~~~~&15022  ~&$15007\pm18\pm24$  ~&15109             ~&$\cdots$           ~&15291           \\
~&$1^{4}D_{\frac{1}{2}^{+}}$      ~&5352   ~&5399(13)           ~&5325              ~&5376               ~&5473              ~~~~~~&14971  ~&$14953\pm17\pm24$  ~&15102             ~&14894              ~&15306           \\
~&$1^{4}D_{\frac{3}{2}^{+}}$      ~&5368   ~&5430(13)           ~&5313              ~&5376               ~&5448              ~~~~~~&14975  ~&$14958\pm17\pm23$  ~&15089             ~&14894              ~&15300           \\
~&$1^{4}D_{\frac{5}{2}^{+}}$      ~&5392   ~&5406(15)           ~&5329              ~&5376               ~&5416              ~~~~~~&14981  ~&$14964\pm17\pm23$  ~&15109             ~&14894              ~&15293           \\
~&$1^{4}D_{\frac{7}{2}^{+}}$      ~&5418   ~&5397(49)           ~&5331              ~&5376               ~&5375              ~~~~~~&14988  ~&$14969\pm16\pm23$  ~&15101             ~&14894              ~&15286           \\
\hline\hline
\end{tabular}}
\end{center}
\end{table*}

There are six $1D$-wave states $|1^4D_{7/2^+,~5/2^+~3/2^+~1/2^+}\rangle$ and $|1^2D_{5/2^+~3/2^+}\rangle$ in
$\Omega_{QQQ}$ spectrum according to the quark model classification (see Table~\ref{wave function table}).
For a comparison, the masses of the 1$D$-wave $\Omega_{ccc}$ and $\Omega_{bbb}$ states predicted in present work together with
those from other works are listed in Table~\ref{2S1D mass table}. Our predictions of the radiative decay properties of the 1$D$-wave states
are also given in Table~\ref{the radiative decay widths}. To our knowledge, no studies of the radiative decay properties of the
1$D$-wave triply heavy baryons can be available in the literature.

\subsubsection{$\Omega_{ccc}(1D)$ states}

The masses for the spin quartets $\Omega_{ccc}(1^4D_{J})$ are predicted to be in the range of $\sim 5.35-5.42$ GeV in present work,
which is compatible with the predictions from Lattice QCD~\cite{ref17} and NRCQM~\cite{ref21}. From
Table~\ref{2S1D mass table}, it is found that the mass order for the spin quartets predicted in the literature
is very different, in this work we predict a normal order, i.e.,
\begin{eqnarray}
M[\Omega_{ccc}(1^4D_{1/2^+})]<M[\Omega_{ccc}(1^4D_{3/2^+})]\ \ \ \ \ \ \ \ \ \ \ \nonumber\\
<M[\Omega_{ccc}(1^4D_{5/2^+})]<M[\Omega_{ccc}(1^4D_{7/2^+})].
\end{eqnarray}
The mass splitting between two adjacent states is about $20$ MeV.

For the spin doublets $\Omega_{ccc}(1^2D_{3/2^+})$ and $\Omega_{ccc}(1^2D_{5/2^+})$, their masses are predicted to be $\sim 5412$ MeV and $\sim 5433$ MeV, respectively, which are compatible with those of Lattice QCD~\cite{ref17} and HCQM~\cite{ref1}.
In this work we find the mass of $\Omega_{ccc}(1^2D_{5/2^+})$ is about 20 MeV
above that of $\Omega_{ccc}(1^2D_{3/2^+})$. The mass order
\begin{eqnarray}
M[\Omega_{ccc}(1^2D_{3/2^+})]<M[\Omega_{ccc}(1^2D_{5/2^+})]
\end{eqnarray}
predicted by us is different from the predictions in Refs.~\cite{ref17, ref1}. To clarify the mass order of these spin multiplets,
the spin dependent interactions should be further studied in future works.

The radiative decay properties of the $1D$-wave $\Omega_{ccc}$ states are
also studied, our results are listed in Table~\ref{the radiative decay widths}.
For the spin quartets $\Omega_{ccc}(1^4D_{J})$, their decay rates into the $1P$-wave
state $\Omega_{ccc}(1P)$ are small. The maximum radiative decay width in the radiative transitions
$\Omega_{ccc}(1^4D_{J})\to \Omega_{ccc}(1P)$ is no more than 1 KeV. However,
the spin doublets $\Omega_{ccc}(1^2D_{3/2^+})$ and $\Omega_{ccc}(1^2D_{5/2^+})$
have large decay rates into the $1P$-wave states via the $E1$ dominated
processes $\Omega_{ccc}(1^2D_{3/2^+})\to \Omega_{ccc}(1P)\gamma$ and $\Omega_{ccc}(1^2D_{5/2^+})\to
\Omega_{ccc}(1^2P_{3/2^-})\gamma$. These radiative decay widths are predicted to be
\begin{eqnarray}
\Gamma[\Omega_{ccc}(1^2D_{3/2^+})\to \Omega_{ccc}(1^2P_{1/2^-})\gamma]&\simeq &107 ~\text{keV},\\
\Gamma[\Omega_{ccc}(1^2D_{3/2^+})\to \Omega_{ccc}(1^2P_{3/2^-})\gamma]&\simeq &34 ~\text{keV},\\
\Gamma[\Omega_{ccc}(1^2D_{5/2^+})\to \Omega_{ccc}(1^2P_{3/2^-})\gamma]&\simeq &122 ~\text{keV}.
\end{eqnarray}
If the spin doublets $\Omega_{ccc}(1^2D_{3/2^+})$ and $\Omega_{ccc}(1^2D_{5/2^+})$ can be
produced in future experiments, these radiative processes may be useful for establishing them.

\subsubsection{$\Omega_{bbb}(1D)$ states}

As shown in Table~\ref{2S1D mass table}, the masses of the $1D$-wave states $\Omega_{bbb}(1D)$ are predicted to
be in the range of $\sim 14.97-15.02$ GeV in present work. The mass order for the six $1D$-wave states is
\begin{eqnarray}
M[\Omega_{bbb}(1^4D_{1/2^+})]<M[\Omega_{bbb}(1^4D_{3/2^+})]\ \ \ \ \ \ \  \ \ \ \nonumber\\
<M[\Omega_{bbb}(1^4D_{5/2^+})]<M[\Omega_{bbb}(1^4D_{7/2^+})]\ \ \ \ \nonumber\\
<M[\Omega_{bbb}(1^2D_{3/2^+})] <M[\Omega_{bbb}(1^2D_{5/2^+})].
\end{eqnarray}
The mass splitting between two adjacent states is about several MeV. It is interesting to find that our prediction of the masses, mass order,
and mass splitting for $1D$-wave states $\Omega_{bbb}(1D)$ are
consistent with those of the Lattice QCD~\cite{ref18}. However, the masses of the $1D$-wave states predicted in this work
are about 100-300 MeV lower than those predicted in Refs.~\cite{ref2,ref1}, while about 100 MeV
higher than those predicted in Ref.~\cite{ref21}. Our predicted mass order for the $1D$-wave states
is also different from those predictions in Refs.~\cite{ref2,ref1}. As a whole there are large uncertainties in the predictions
of the mass spectrum of the $1D$-wave states $\Omega_{bbb}(1D)$, more theoretical studies are needed.

In present work we also give our predictions of the radiative decays of $1D$-wave states $\Omega_{bbb}(1D)$.
Our results are collected in Table~\ref{the radiative decay widths}.
From the table it is found that for the spin quartets $\Omega_{ccc}(1^4D_{J})$, their partial widths of the
radiative decays into the $1P$-wave states $\Omega_{bbb}(1P)$ are tiny ($\mathcal{O}(1)$ eV)
for the absences of the $E1$ transitions. However, for spin doublets $\Omega_{bbb}(1^2D_{3/2^+})$ and
$\Omega_{bbb}(1^2D_{5/2^+})$, there are fairly large decay rates into the $1P$-wave
states $\Omega_{bbb}(1P)$ via the $E1$ dominant transitions.
These radiative decay widths are predicted to be
\begin{eqnarray}
\Gamma[\Omega_{bbb}(1^2D_{3/2^+})\to \Omega_{bbb}(1^2P_{1/2^-})\gamma]&\simeq & 7.8 ~\text{keV},\\
\Gamma[\Omega_{bbb}(1^2D_{3/2^+})\to \Omega_{bbb}(1^2P_{3/2^-})\gamma]&\simeq &2.8 ~\text{keV},\\
\Gamma[\Omega_{bbb}(1^2D_{5/2^+})\to \Omega_{bbb}(1^2P_{3/2^-})\gamma]&\simeq &8.4 ~\text{keV}.
\end{eqnarray}
These partial widths for the $1D$-wave $\Omega_{bbb}$ states are about one order of
magnitude smaller than those corresponding processes of the $1D$-wave $\Omega_{ccc}$ states.

\subsection{$2S$ states}

There are two $2S$-wave $\Omega_{ccc}$/$\Omega_{bbb}$ states with $J^P=1/2^+$ and $J^P=3/2^+$ according to the quark model classification (see Table~\ref{wave function table}). For a comparison, our predicted masses of the 2$S$-wave $\Omega_{ccc}$ and $\Omega_{bbb}$ states together with those of other theoretical predictions have been listed in Table~\ref{2S1D mass table}. Furthermore, our predictions of the radiative decay properties of the $2S$-wave $\Omega_{ccc}$ and $\Omega_{bbb}$ states are given in Table~\ref{the radiative decay widths}. To our knowledge, no studies of the radiative decay properties of the 2$S$-wave triply heavy baryons can be available in the literature.

\subsubsection{$\Omega_{ccc}(2S)$ states}

Our predicted masses for the $2S$-wave $\Omega_{ccc}$ states $\Omega_{ccc}(2^2S_{1/2^+})$ and $\Omega_{ccc}(2^4S_{3/2^+})$
are $\sim 5373$ MeV and $\sim 5285$ MeV, respectively, which are compatible with the Lattice
QCD predictions in Ref.~\cite{ref17}. The mass splitting between these two $2S$-wave state predicted in present work, $\sim 90$ MeV,
is also in good agreement with that of the Lattice QCD~\cite{ref17}. It should be mention that there are only a few predictions of masses of the $2S$-wave $\Omega_{ccc}$ states. The mass range predicted in this work roughly agrees with the other quark
model predictions~\cite{ref2,ref21,ref1}, although the predicted mass splitting between the two $2S$-wave
$\Omega_{ccc}$ states is different with each other.

The radiative decay rates of $\Omega_{ccc}(2^2S_{1/2^+})$
into $\Omega_{ccc}(1^2P_{1/2^-})\gamma$ and $\Omega_{ccc}(1^2P_{3/2^-})\gamma$ final states are relatively large.
The radiative partial widths are predicted to be
\begin{eqnarray}
\Gamma[\Omega_{ccc}(2^2S_{1/2^+})\to \Omega_{ccc}(1^2P_{1/2^-})\gamma]\simeq 20 ~\text{keV},\\
\Gamma[\Omega_{ccc}(2^2S_{1/2^+})\to \Omega_{ccc}(1^2P_{3/2^-})\gamma]\simeq 27 ~\text{keV}.
\end{eqnarray}
These radiative processes may be useful for establishing the $2S$-wave $\Omega_{ccc}$ states in future experiments.
However, the radiative decay widths of $\Omega_{ccc}(2^4S_{3/2^+})$ into the $1P$-wave states $\Omega_{bbb}(1P)$ are tiny ($\mathcal{O}(1)$ eV)
for the absences of the $E1$ transitions.

\subsubsection{$\Omega_{bbb}(2S)$ states}

Our predicted masses for the $2S$-wave $\Omega_{bbb}$ states $\Omega_{bbb}(2^2S_{1/2^+})$ and $\Omega_{bbb}(2^4S_{3/2^+})$
are $\sim 14959$ MeV and $\sim 14848$ MeV, respectively, which are compatible with the Lattice
QCD predictions in Ref.~\cite{ref18}. The mass splitting between these two $2S$-wave state predicted in present work, $\sim 110$ MeV,
is also in good agreement with that of the Lattice QCD~\cite{ref18}. Our predicted mass of $\Omega_{bbb}(2^4S_{3/2^+})$ is also
close to recent quark model prediction, 14805 MeV, in Ref.~\cite{ref21}.  However, our predictions of the masses
for these $2S$-wave $\Omega_{bbb}$ states are about 200-300 MeV lower than the other quark model predictions in Refs.~\cite{ref2,ref1}.

The $E1$ dominant radiative transitions of $\Omega_{bbb}(2^2S_{1/2^+})\to \Omega_{bbb}(1P)\gamma$ might play a crucial
role in its decays. The radiative partial widths for these processes are predicted to be
\begin{eqnarray}
\Gamma[\Omega_{bbb}(2^2S_{1/2^+})\to \Omega(1^2P_{1/2^-})\gamma]= 0.99 ~\text{keV},\\
\Gamma[\Omega_{bbb}(2^2S_{1/2^+})\to \Omega(1^2P_{3/2^-})\gamma]= 1.46 ~\text{keV}.
\end{eqnarray}
Compared them with the partial widths of $\Omega_{ccc}(2^2S_{1/2^+})\to \Omega_{ccc}(1^2P_{1/2^-})\gamma,
\Omega_{ccc}(1^2P_{3/2^-})\gamma$, it is found that
\begin{eqnarray}
\mathcal{R}=\frac{\Gamma[\Omega_{bbb}(2^2S_{1/2^+})\to \Omega_{bbb}(1P_J)\gamma]}{\Gamma[\Omega_{ccc}(2^2S_{1/2^+})\to \Omega_{ccc}(1P_J)\gamma]}\simeq \frac{1}{20}.
\end{eqnarray}
It indicates that the $\Omega_{bbb}(2^2S_{1/2^+})$ state may be more difficultly observed
than $\Omega_{ccc}(2^2S_{1/2^+})$ via the radiative decay processes. Finally,
it should be mentioned that the radiative decay rates of $\Omega_{bbb}(2^4S_{3/2^+})$
into the $1P$-wave states $\Omega_{bbb}(1P)$ are tiny for the absences of the $E1$ transitions.
The predicted partial widths are less than 1 eV. Thus, the radiative decay processes of $\Omega_{bbb}(2^4S_{3/2^+})$
might be less helpful for establishing it in experiments.

\section{Summary}\label{Summary}

In this work, we calculate the $\Omega_{ccc}$ and $\Omega_{bbb}$ spectrum up to the $N=2$ shell within a potential model.
The potentials are determined by fitting the mass spectra of charmonium and bottomonium in our previous works.
For the ground states $\Omega_{ccc}$ and $\Omega_{bbb}$, our predicted masses are
$\sim 4828$ MeV and $\sim 14432$ MeV, respectively. Compared with the results of the lattice QCD, it is found
our predicted mass for $\Omega_{ccc}$ just lies the upper limit of the predictions in Refs.~\cite{ref12,ref14,ref15},
while our predicted mass for $\Omega_{bbb}$ is about 60 MeV above the predictions in Refs.~\cite{ref12,ref18}.
Furthermore, our predictions of the mass ranges for the $1P$-, $1D$-, and $2S$-wave excited
$\Omega_{ccc}$ and $\Omega_{bbb}$ states are in good agreement with the Lattice QCD predictions~\cite{ref17,ref18}.
It should be pointed out that mass orders for the spin multiplets in the excited
$\Omega_{ccc}$ and $\Omega_{bbb}$ states predicted in the literature is very different.
To clarify the mass order of these spin multiplets, the spin dependent integrations should be further studied in future works.

Moreover, by using the predicted masses and wave functions from the potential model,
the radiative transitions for the $1P\to 1S$, $1D\to 1P$, and $2S\to 1P$
are evaluated for the first time with a constituent quark model. For the $\Omega_{ccc}$ sector,
the transition rates for $1P \to 1S$ might be sizeable, the partial widths are about
several keV; the transition rates for the $E1$ dominant decay processes
$1^2D_{3/2^+}\to 1^2P_{1/2^+,3/2^+}$, $1^2D_{5/2^+}\to 1^2P_{3/2^+}$,
$2^2S_{1/2^+}\to 1^2P_{1/2^+,3/2^+}$ are relatively large, their partial widths are
predicted to be about 10s keV. For the $\Omega_{bbb}$ sector, the
partial widths of the corresponding transitions mentioned above are about one or two
order of magnitude smaller than those for the $\Omega_{ccc}$ sector.
To better understand the radiative decay properties of the excited $\Omega_{ccc}$ and $\Omega_{bbb}$
baryon states, more studies are hoped to be carried out in theory.

\section*{Acknowledgement}

This work is supported by the National Natural Science Foundation of China under Grants No.~11775078, No.~U1832173, and No.~11705056.

\bibliographystyle{unsrt}

\end{document}